\documentclass[twocolumn,preprintnumbers,amsmath,amssymb]{revtex4}
\usepackage{graphicx}

\usepackage{amsmath}
\usepackage{amssymb}
\usepackage{amsthm}
\usepackage{amsfonts}
\usepackage{enumerate}

\usepackage{centernot}
\usepackage{tikz}

\usepackage{subfigure}

\begin{document}

\title{Magneto-optical Conductance of Kane Fermion Gas in Low Frequencies}

\author{Xi Luo$^1$}
\thanks{These two authors contribute equally.}
\author{ Yu-Ge Chen$^{2,3}$}
\thanks{These two authors contribute equally.}
\author{ Yue Yu$^{2,3,4}$}
\email{yyu@itp.ac.cn}
\affiliation {1. Department of physics, College of Science, University of Shanghai for Science and Technology, Shanghai 200093, China\\
	2.Center for Field
	Theory and Particle Physics, Department of Physics, Fudan University, Shanghai 200433,
	China \\
	3. State Key Laboratory of Surface Physics, Fudan University, Shanghai 200433,
	China\\
	4. Collaborative Innovation Center of Advanced Microstructures, Nanjing 210093, China}

\begin{abstract}
	Kane fermion is the counterpart of the Dirac fermion with pseudospin-1. Due to the existence of a bunch of gapless modes associated with  Landau levels, the magnetic transport property of Kane fermion gas is very different from that of the Dirac  semimetal.  We calculate the magneto-optical conductance of the Kane fermion gas. We find that these gapless modes will contribute to  a series of resonant peaks in low frequencies. We find that these peaks can explain the low frequency absorbance spectrum in a recent experiment for the Kane fermion material Hg$_{1-x}$ Cd$_x$Te.
\end{abstract}

\date{\today}

\maketitle

\section{introduction}

The discoveries of Weyl and Dirac semimetals \cite{weyl1,weyl2,weyl3,dirac1,dirac2,dirac3,dirac4} open a gate to observe new topological phenomena in condensed matter systems, such as the Fermi arcs, giant negative magneto-resistance, and three dimensional quantum anomalous Hall effect \cite{burkov2014}. These exotic properties are rooted from the non-trivial Berry's phase of the Weyl nodes in spectrum: The linearly touched Weyl nodes serve as monopoles of the Berry's curvature \cite{weyl1,weyl-rmp}. One of the evidences of linear band touching structure of Weyl and Dirac semimetals comes from the measurement of its magneto-optical conductance, i.e., the resonant frequency peaks between the extrema of Landau levels are proportional to the square root of the external magnetic field \cite{linear1,linear2,linear3}.  

Besides the Weyl and Dirac semimetals, new kinds of fermions have been proposed and observed in condensed matter systems \cite{kane, mo-kane2, type2,new,hourglass}. The triply degenerate node in the band structure of MoP is an example that does not have an analog in high energy systems \cite{tri}.  Such a fermionic quasiparticle emerges because these "relativistic" quasiparticles in condensed matter systems are not confined by the  "Lorentz symmetry".  A more interesting triply degenerate fermion might be the photon-like fermions, the pseudospin-1 generalization of Dirac and Weyl fermions. The former is known as Kane fermion and has been recently experimentally observed in Hg$_{1-x}$ Cd$_x$Te \cite{kane} and Cd$_3$As$_2$ \cite{mo-kane2}. The latter is the spinless courterparter of the Kane fermion and is proposed to exist in materials with space group 199 and 214, such as Pd$_3$Bi$_2$S$_2$ and Ag$_3$Se$_2$Au \cite{new}. Recently, APd$_3$(A=Pb, Sn) \cite{tri2}, LaPtBi \cite{tri3}, and ZrTe \cite{tri4} are also proposed to have triple nodal points. More possible materials would be found by the method of symmetry indicators \cite{sym1,sym2,sym3}. 

{As well known, a stable Dirac node against the perturbation needs to be protected by an extra symmetry \cite{extra1,extra2}. However, in theoretical study, a gapless Dirac equation can be decomposed into two decoupled Weyl equations with opposite chirality.  Similarly, a gapless Kane fermion can be decomposed into two three component photon-like fermions.  Furthermore,  we will see that due to a quasi-one dimensional particle-hole symmetry of the gapless Kane fermion model,  the magneto-optical conductance of the gapless Kane fermion gas is simply twice as that of the photon-like gas. Therefore, we focus first on the photon-like fermion. }

Although the topological properties of the photon-like fermion are similar to those of Weyl fermion, e.g., the monopole charge of the photon-like fermion is twice as large as that of the Weyl fermion, there is a longitudinal photon-like flat band in a free photon-like fermion gas which is absent in Weyl semimetal. After applying an external magnetic field, this flat band becomes a bunch of gapless modes in Landau levels, and they generate anomalous magnetic transport properties for the photon-like fermion gas. The anomalous magnetic resistance and  extra quantum oscillation of the density of states of the photon-like fermion gas were studied \cite{photon-like fermion}.

The magneto-optical measurement {is another physical observable, and explores the effects of the Landau levels, both the ordinary gapped Landau levels and the extra gapless ones.} In this paper, we will {study} the magneto-optical conductance of {Kane} fermion gas. We first calculate the magneto-optical conductance contributed by a single photon-like fermion node.  We find that there are intriguing resonant peaks near the zero frequency in the magneto-optical conductance. These peaks are originated from the transition between the gapless Landau modes. Furthermore,
{the peaks stemming from the resonance between the gapless Landau modes and the ordinary gapped ones will split, and the splitting is proportional to the band width of the gapless modes,} which won't show up if the gapless Landau modes are exactly flat. Because of an emergent quasi-one dimensional particle-hole symmetry of the Hamiltonian, the contribution to the magneto-conductance from the photon-like fermion node with opposite chirality is the same as its chiral partner's. Therefore, the magneto-optical conductance in the {Kane} fermion gas can be obtained  by totting-up that from each single photon-like fermion node. We also discuss the linear dependence between the resonant frequency $\omega$ and the square root external magnetic field $B$, which indicates the linear dispersion behavior of the gapless Landau modes.

 The Kane fermion, has been recently experimentally observed in Hg$_{1-x}$ Cd$_x$Te \cite{kane}. The experimentally measured high frequency magneto-optical resonance is consistent with the characteristic of the massless Kane fermion.  Recent experiment confirms these magneto-optical signature of massless Kane fermions in Cd$_3$As$_2$ \cite{mo-kane2}.  Numerically, the magneto-optics of the Kane fermion was studied \cite{mo-kane3,mo-kane}.  However, for the gapless Kane fermion, the contributions from the gapless Landau modes to the magneto-optical conductance were not considered in the numerical calculations. Experimentally, the measurement for Cd$_3$As$_2$ does reach that low frequency while the data for Hg$_{1-x}$ Cd$_x$Te was well explained. On the other hand, as we have mentioned,  the magneto-optical conductance of the gapless Kane fermion gas is double as that of the photon-like gas. Thus, our study can be directly applied to the gapless Kane fermion. Especially, we compare our calculation for the magneto-optical conductance near the zero frequency with the data for Hg$_{1-x}$ Cd$_x$Te in \cite{kane} and give a reasonable explanation to the low frequency peaks in the magneto-optical absorbance.

This paper was organized as follows: In Sec. II, we calculate the magneto-optical conductance of the photon-like fermion gas. {We also show the dependence of chemical potential and temperature of the magneto-optical conductance.} 
In Sec. III, {We study the gapless Kane fermion model and compare the numerical results with experimental data. The effect of weak disorder on the low energy states of the Kane fermion is also discussed} In Sec. IV, we introduce a quadratic correction term to the spectrum of the photon-like fermion {as well as} the magneto-optical conductance. The last section is devoted to conclusions.    

\section{magneto-optical conductance of photon-like fermion gas}

\subsection{Single photon-like fermion node}

        
\begin{figure}[ptb]
\centering
\includegraphics[width=0.3\textwidth]{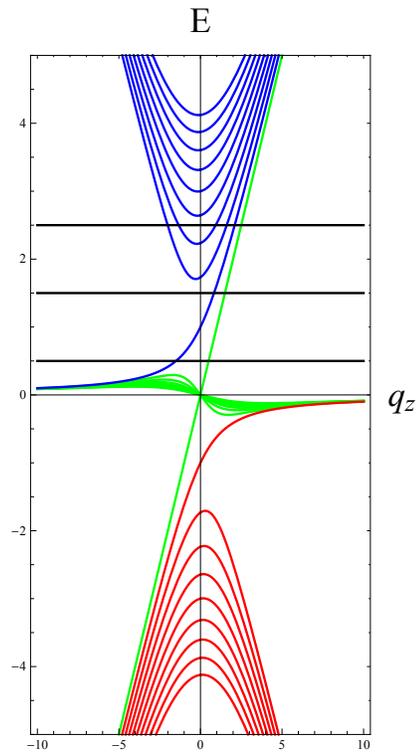}

\caption{ (color online) A sketch diagram of Landau Level structure for photon-like fermion. {We choose $l_B=1$ in Eq. (\ref{hb}). The green curves are the chiral lowest Landau level and the gapless ones, the blue curves are the 1st Landau level and the gapped ones with positive energies, and the red curves are the 1st Landau level and the gapped ones with negative energies. The black lines correspond to the chemical potential used in Fig. \ref{xx-xy}(a).}}
\label{fig1}
\end{figure}

Under a magnetic field ${\bf B}=\nabla\times {\bf A}=-B\hat{z}$ in the $z$ direction, the momentum operator is given by ${\bf D}=-i\hbar \nabla+\frac{e}c{\bf A}$.  The Hamiltonian  near the photon-like fermion node  reads \cite{photon-like fermion}
\begin{equation}
H_{B}=\hbar v_F
\left(
\begin{array}{ccc}
p_z
 & a^\dag/l_B
 & 0 \\ 
a/l_B
 & 0
 & a^\dagger/l_B \\ 
 0
 & a/l_B
 & -p_z
\end{array}
\right), \label{hb}
\end{equation}
with {the Fermi velocity $v_F$, the Bloch wave vector $p_i$, the annihilation operator of Landau levels $a=l_B(D_x+iD_y)/\sqrt{2}$, and the magnetic length $l_B=\sqrt{\hbar /eB}$ \cite{kane}. In usual materials, the Fermi velocity is of the order $10^6m/s$, the strength of the magnetic field is of the order of $10T$, therefore the magnetic length $l_B\sim 10^{-8}m$. The gap between adjacent Landau levels is of the order $\bar v_F/l_B\sim 100meV$, which is experimentally accessible for magneto-optical measurements \cite{kane}. For later convenience, we choose $\hbar=k_B=v_F=e=1$ except explicitly shown.} The eigen wave function $\Psi(n)$ for the $n$th Landau level has the form $(\alpha_{n}\phi_{n},\beta_{n-1}\phi_{n-1},\gamma_{n-2}\phi_{n-2})^T$, where $\phi_n$ is the $n$th normalized Landau level wave function {with respect to $a^\dag a$} and $\alpha,\beta,\gamma$ are the coefficients. There are two special cases. $(\phi_0,0,0)^T$ and $(\alpha_1 \phi_1,\beta_0\phi_0,0)^T$ correspond to the zeroth and the first Landau levels. We plot several Landau level structures in Fig. \ref{fig1}.

The zeroth and the first Landau levels have the dispersions $E_0=\hbar v_Fp_z$, and $E_1=\hbar v_F p_z\pm\sqrt{p_z^2+4B/l^2_B}$. The robustness of these two Landau levels is guaranteed by the chiral anomaly. For the Landau level with $n\geq2$, there are three eigenvalues \cite{photon-like fermion}
\begin{eqnarray}
	E_n^+&=&\hbar v_F[(p_z/2+\sqrt{\Delta})^{\frac{1}{3}}+(p_z/2-\sqrt{\Delta})^{\frac{1}{3}}],\nonumber\\
	E_n^-&=&\hbar v_F[\omega_+(p_z/2+\sqrt{\Delta})^{\frac{1}{3}}+\omega_-(p_z/2-\sqrt{\Delta})^{\frac{1}{3}}], \label{en}\\
	E_n^0&=&\hbar v_F[\omega_-(p_z/2+\sqrt{\Delta})^{\frac{1}{3}}+\omega_+(p_z/2-\sqrt{\Delta})^{\frac{1}{3}}],\nonumber
\end{eqnarray}
where $\omega_\pm=(-1\pm\sqrt{3}i)/2$ and $\Delta=p_z^2/4-((2n-1)*l_B^2+p_z^2)^3/27$. {Since $\Delta<0$, $E_n^*=E_n$, namely, all the energies (\ref{en}) are real.} In the $p_z\rightarrow 0$ limit, 
\begin{eqnarray}
E_n^\pm\rightarrow \pm \hbar v_F l_B^{-1}\sqrt{2n-1},\label{low}
\end{eqnarray}
 and $E_n^0=0$ which reduce to the two dimensional results \cite{lieb}. 


$E_n^\pm$ can be thought as a pseudo-spin $1$ {counterpart} of the Landau levels of  Weyl fermion while  $E_n^0$ emerge from the zero field flat band when the external magnetic field is applied. These modes are absent for  Weyl fermion \cite{photon-like fermion}. For a small $p_z$, 
$E_n^0\rightarrow -\frac{\hbar v_Fp_z}{2n-1}$ and when $|p_z|\rightarrow \infty$, $E_n^0\rightarrow 0$. These behaviors suggest that the $E_{n>1}^0(p_z)$ are gapless modes near {$p_z=0$  associated with} the Landau level $n>1$ (shorten as the gapless Landau modes). {Near $p_z=0$,} these gapless Landau modes have opposite chirality to that of the zeroth Landau level (see Fig. \ref{fig1}).

\begin{figure}
	\begin{minipage}{0.96\linewidth}
		\centerline{\includegraphics[width=1\textwidth]{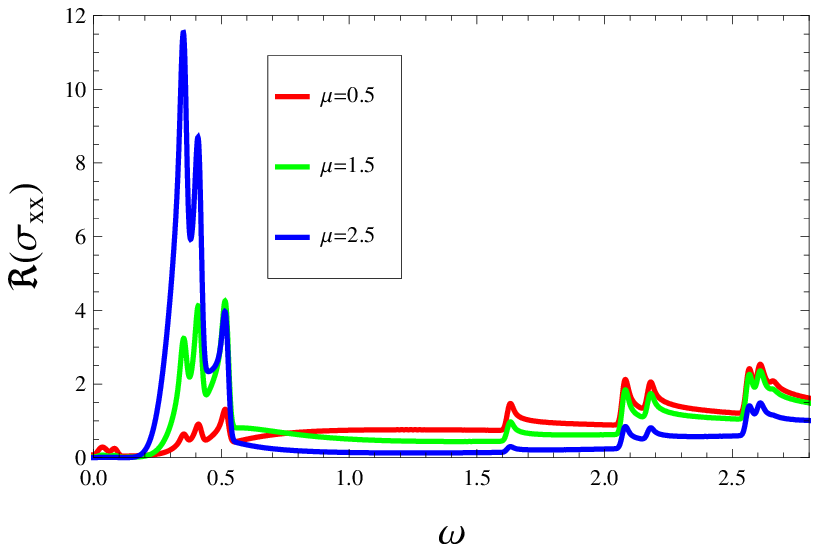}}
		\centerline{(a)}
	\end{minipage}\\
	\begin{minipage}{0.96\linewidth}
		\centerline{\includegraphics[width=1\textwidth]{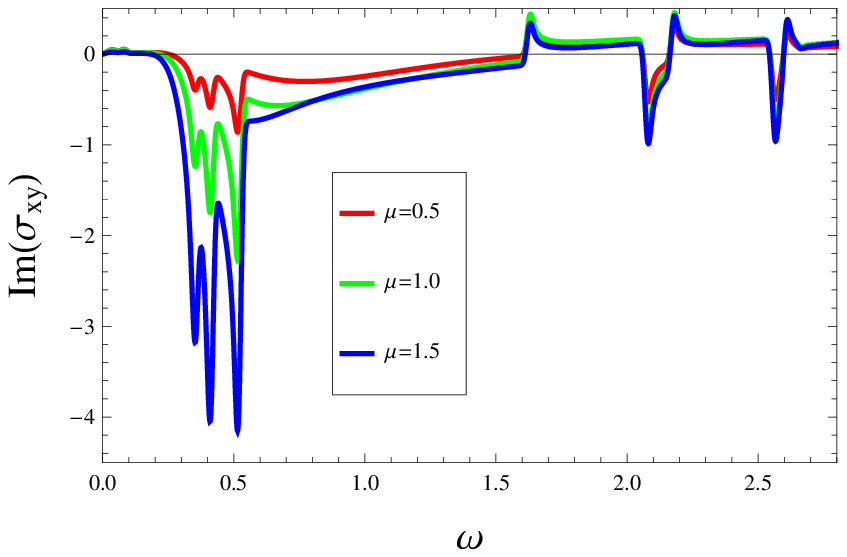}}
		\centerline{(b)}
	\end{minipage}
	\caption{ (color online) Magneto-optical conductivity of the phone-like fermion.   (a) is the real part of $\sigma_{xx}$ with{ $\mu=0.5/l_B$, $\mu=1.5/l_B$ and $\mu=2.5/l_B$ }respectively. (b) is the imaginary part of {$\sigma_{xy}$ with $\mu=0.5/l_B$, $\mu=1.0/l_B$ and $\mu=1.5/l_B$} respectively. 
		}
	\label{xx-xy}
\end{figure}

In a previous work, {we studied the magnetic transport properties and the quantum oscillations of the density of states of the photon-like fermion gas, especially, the effects of the gapless Landau modes \cite{photon-like fermion}. The magneto-optical measurement is another possible way to explore the Landau level physics as well as the novel gapless modes of the photon-like fermion gas.} Using the Kubo formula, we define the magneto-optical current 
	\begin{equation}
	J_\alpha=\sigma_{\alpha \beta}E_\beta,
	\end{equation}
	where the magneto-optical conductivity tensor in the Landau level representation is given by \cite{mahan},   	
\begin{eqnarray}
\sigma_{\alpha\beta}(\omega)&=&\frac{-i}{2\pi l_B^2}\sum_{n,n',s,s'}\int\frac{dp_z}{2\pi}\frac{f(E_n^s)-f(E_{n'}^{s'})}{E_n^s-E_{n'}^{s'}}\nonumber\\
&&\times\frac{\langle \Psi_{ns}|j_\alpha|\Psi_{n's'}\rangle\langle \Psi_{ns}|j_\beta|\Psi_{n's'}\rangle}{ \omega +E_{ns}-E_{n's'}+i0^+},\label{mofer}
\end{eqnarray}
where
\begin{equation}
f(E_{ns})=\frac{1}{\exp(\frac{E_{ns}-\mu}{T})+1},
\end{equation}
is the Fermi distribution. $n$ is the Landau level index. $s=0,\pm$ are the band indices. Furthermore, the current matrices are $j_\alpha=\partial H/\partial p_\alpha$, namely,
\begin{equation}
j_x=\frac{1}{\sqrt{2}}
\left(
\begin{array}{ccc}
0
& 1
& 0 \\ 
1
& 0
& 1 \\ 
0
& 1
& 0
\end{array}
\right), \quad
j_y=\frac{1}{\sqrt{2}}
\left(
\begin{array}{ccc}
0
& -i
& 0 \\ 
i
& 0
& -i \\ 
0
& i
& 0
\end{array}
\right). 
\end{equation}
The expressions of the currents $j_x$ and $j_y$  gives the selection rule $n \rightarrow n\pm1$ for a given $n$th Landau level. We plot the absorptive part of the magneto-optical conductance, namely, the real part of $\sigma_{xx/yy}$ and the imaginary part of $\sigma_{xy}$ in Fig. \ref{xx-xy}. {In the numerical calculation, we replace the delta function by a Gaussian distribution with a standard deviation of $\sigma=0.01$.}

\begin{figure}
	\begin{minipage}{0.48\linewidth}
		\centerline{\includegraphics[width=1\textwidth]{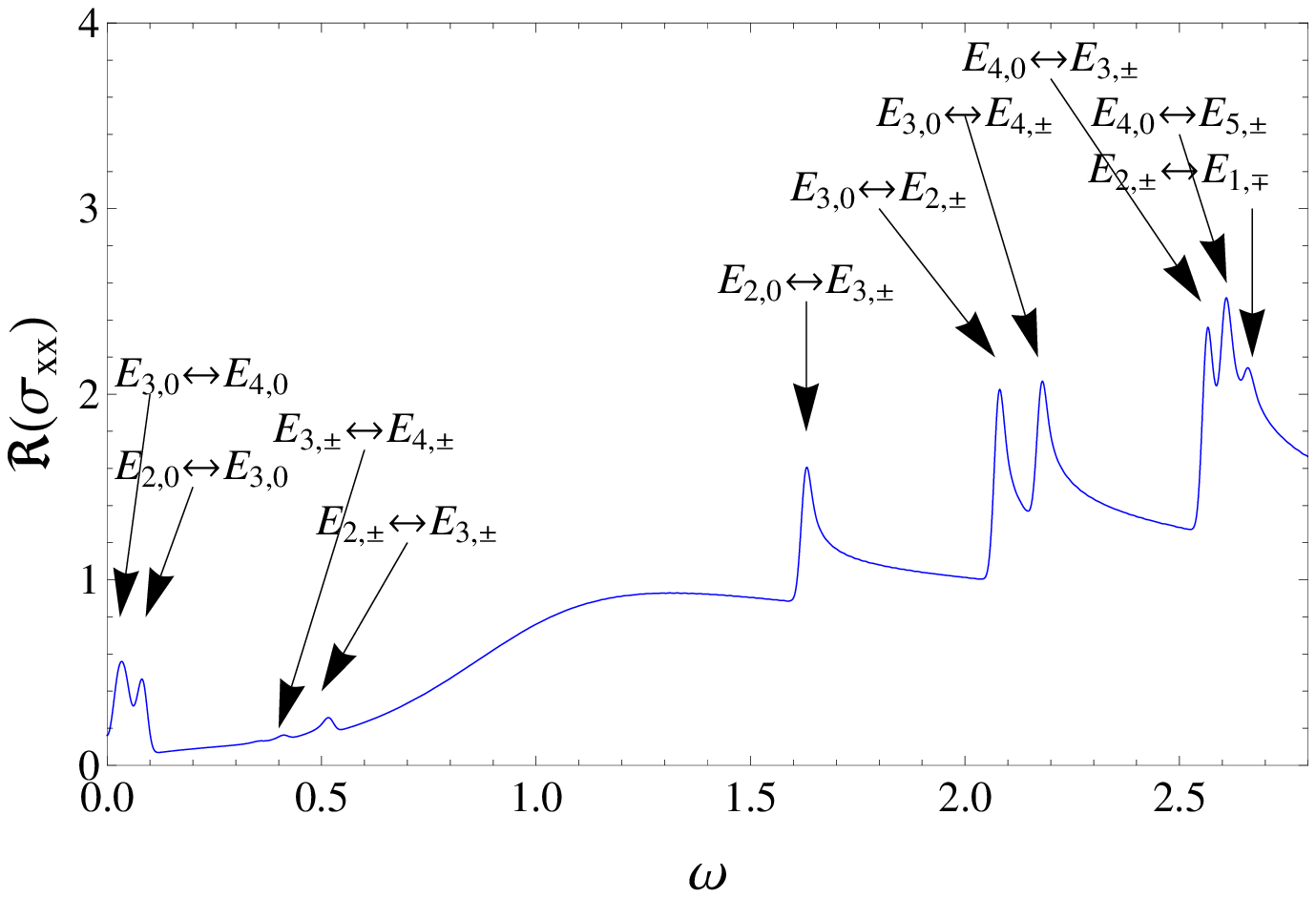}}
		\centerline{(a)}
	\end{minipage}
	\begin{minipage}{0.48\linewidth}
		\centerline{\includegraphics[width=1\textwidth]{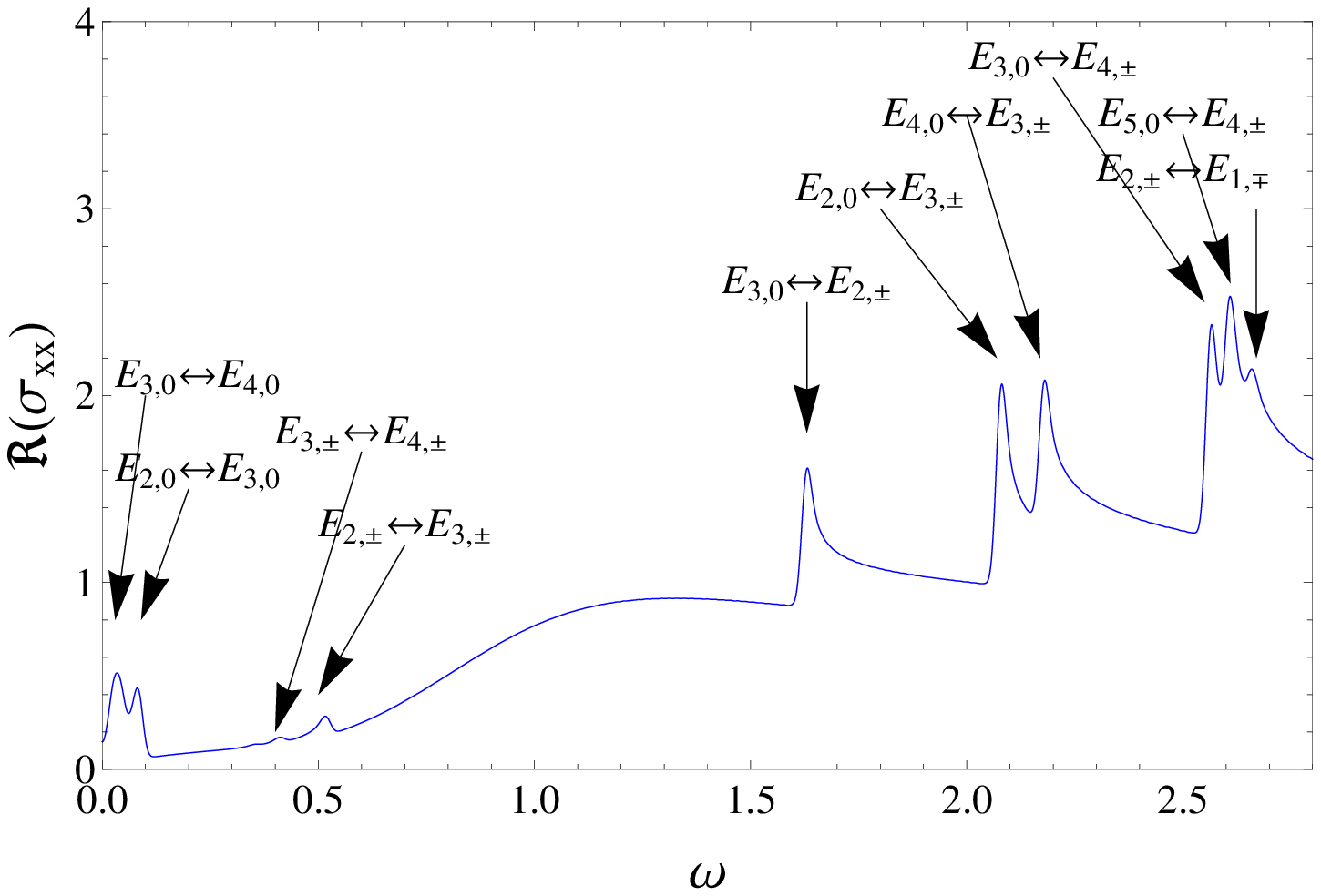}}
		\centerline{(b)}
	\end{minipage}\\
	\begin{minipage}{0.48\linewidth}
		\centerline{\includegraphics[width=1\textwidth]{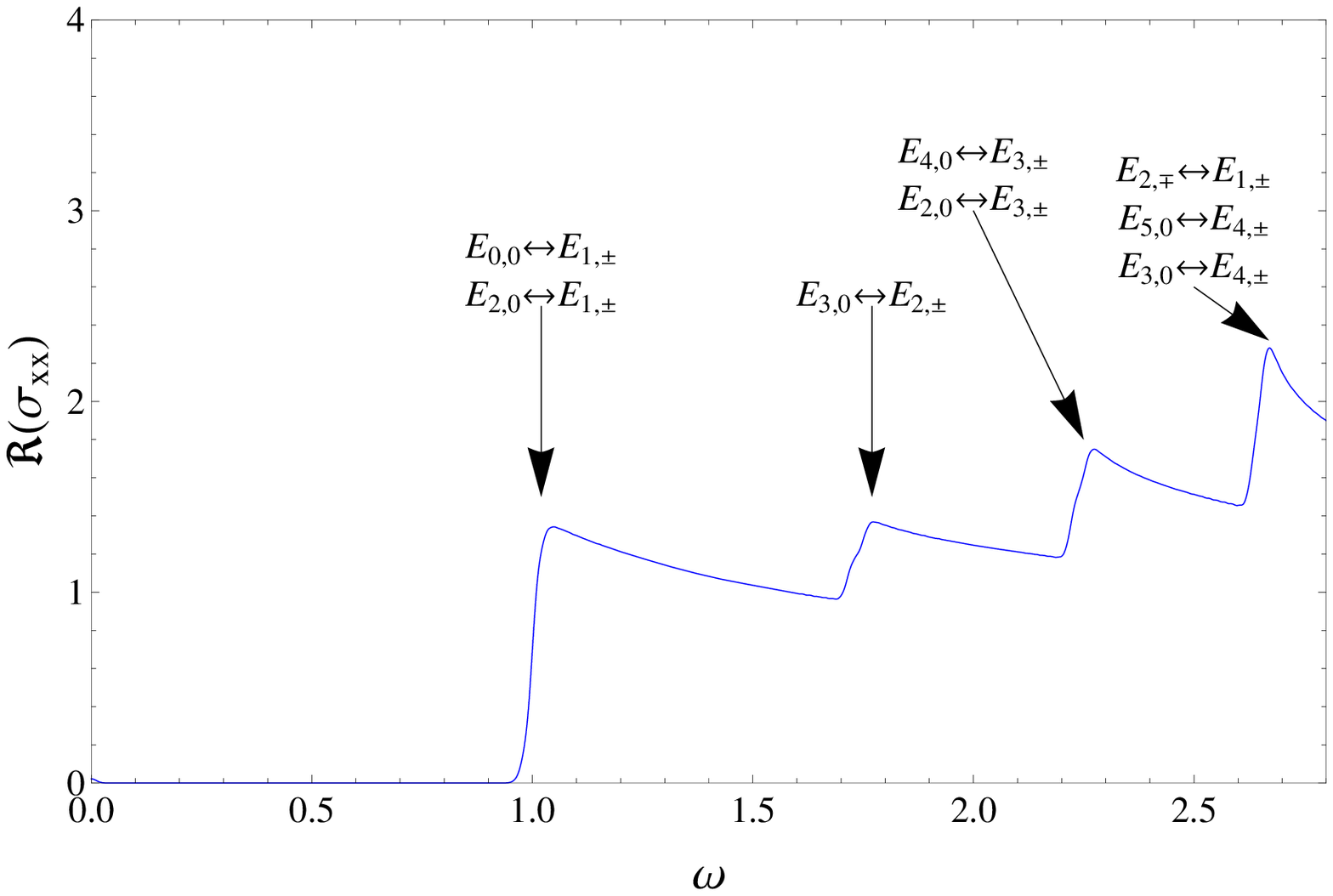}}
		\centerline{(c)}
	\end{minipage}
	\begin{minipage}{0.48\linewidth}
		\centerline{\includegraphics[width=1\textwidth]{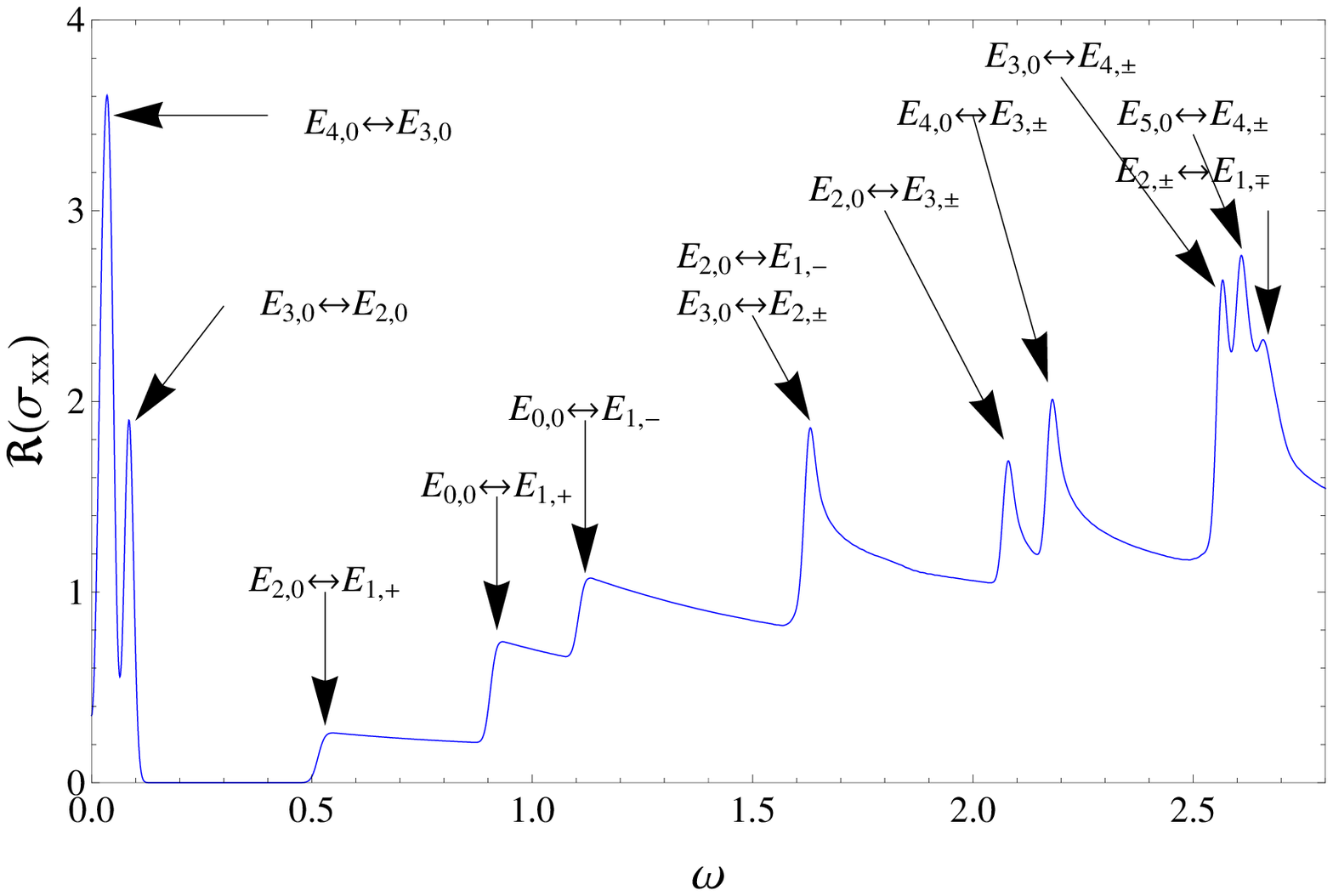}}
		\centerline{(d)}
	\end{minipage}
	\caption{ (color online) {The dependence on temperature and chemical potential of the real part of $\sigma_{xx}$ are plotted in (a) with $T=0.3/l_B$, $\mu=0.01/l_B$, (b) with $T=0.3/l_B$, $\mu=0.2/l_B$, (c) with $T=0.0015/l_B$, $\mu=0.01/l_B$, and (d) with $T=0.0015/l_B$, $\mu=0.2/l_B$. The origin of each peak is also shown. }
	}
	\label{xx}
\end{figure}

In terms of the numerical calculation, we notice that the resonant frequencies in Re$\sigma_{xx/yy}$ and Im$\sigma_{xy}$ appear as {extrema} of differences between the Landau levels respecting the selection rules {in the high temperature regime (see Fig. \ref{xx-xy}(a) and \ref{xx-xy}(b)). While in the low temperature regime, the peaks appear as the difference between the Landau levels near $p_z=0$ because of the strong suppression by the Fermi-Dirac distribution function (see Fig. \ref{xx}(c) and \ref{xx}(d)). Take the curve in Fig \ref{xx}(a) as an example in the numerical calculation where $T=0.3/l_B$ and $\mu=0.01/l_B$ (for $B\sim 10T$, $T\sim 300K$). } We consider the contributions from the $0$-th Landau level to the $5$-th one. In this case, there are 10 peaks in the diagram counting from the zero frequency. The first peak near the zero frequency stems from $E_{4,0}\leftrightarrow E_{5,0}$ and $E_{3,0} \leftrightarrow E_{4,0}$, where the notation ``$E_{4,0}\leftrightarrow E_{5,0}$'' stands for the transition between the $E_{4,0}$ and $E_{5,0}$ bands. Actually there are two peaks, but their energy differences are too small to be identified. The second peak from the zero frequency stems from $E_{2,0}\leftrightarrow E_{3,0}$ bands. If we take more bands into account, there will be almost continuous absorbing frequencies near zero frequency, which indicates the existence of the $E_{n,0}$ bands, and {the differences of the extrema of their energies} are small. The $3$-rd, and $4$-th peaks around {$\omega=0.4/l_B$} come from $E_{4,\pm} \leftrightarrow E_{3,\pm}$, and $E_{3,\pm} \leftrightarrow E_{2,\pm}$ bands, similar to the case of ordinary gapped Landau levels. The first five peaks with frequency larger than {$\omega=1.5/l_B$} come from the transition between $E_{n,+}$ and $E_{n\pm1,0}$ bands, or $E_{n\pm1,0}$ and $E_{n,-}$. For instance, the peak near {$\omega=1.6/l_B$} corresponds to $E_{2,+}\leftrightarrow E_{3,0}$ and $E_{3,0}\leftrightarrow E_{2,-}$. The last peak comes from $E_{1,+}\leftrightarrow E_{2,-}$ and $E_{2,+}\leftrightarrow E_{1,-}$. For higher frequencies which are not shown in Fig. \ref{xx}(a), there will be transitions between $E_{n,-}$ and $E_{n\pm1,+}$ bands, as expected. 
 
{The main differences in the low temperature regime are the strong chemical potential dependent behavior and the existence of the $E_{0,0}\leftrightarrow E_{1,\pm}$ and $E_{2,0}\leftrightarrow E_{1\pm}$ peaks. These peaks exist because of the strong influence from the Fermi-Dirac distribution at low temperature. In this case, the resonant peaks raise from the energy difference between allowed Landau levels near $p_z=0$. In the $p_z=0$ limit, our results differ from the Hg$_{1-x}$ Cd$_x$Te only by a Zeeman term \cite{kane}.  In Fig. \ref{xx}(c) and (d), we choose $T=0.0015/l_B$ ($B\sim 10T$, $T\sim 1K$), and $\mu=0.01/l_B$ and $0.2/l_B$ respectively. The resonance amplitude becomes stronger when the chemical potential $\mu$ gets closer to the transition Landau levels, and vice versa, such as the two peaks near $\omega=0.1/l_B$. For a larger chemical potential, the resonant peaks will split, and the splitting is proportional to the difference between the maximums of $E_{n-1,0}$ and $E_{n+1,0}$, roughly, the band width of the gapless modes. The peak splitting due to increasing the chemical potential at low temperature is another significant evidence for the existence of the gapless Landau levels instead of the exact flat Landau level bands.}
%
%
%
%

\begin{figure}
	\begin{minipage}{0.49\linewidth}
		\centerline{\includegraphics[width=1\textwidth]{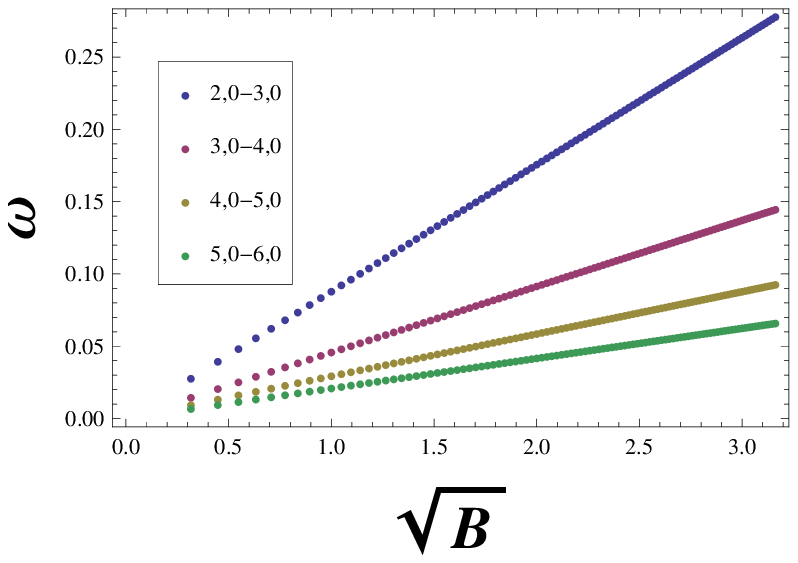}}
		\centerline{(a)}
	\end{minipage}
	\hfill
	\begin{minipage}{.49\linewidth}
		\centerline{\includegraphics[width=1\textwidth]{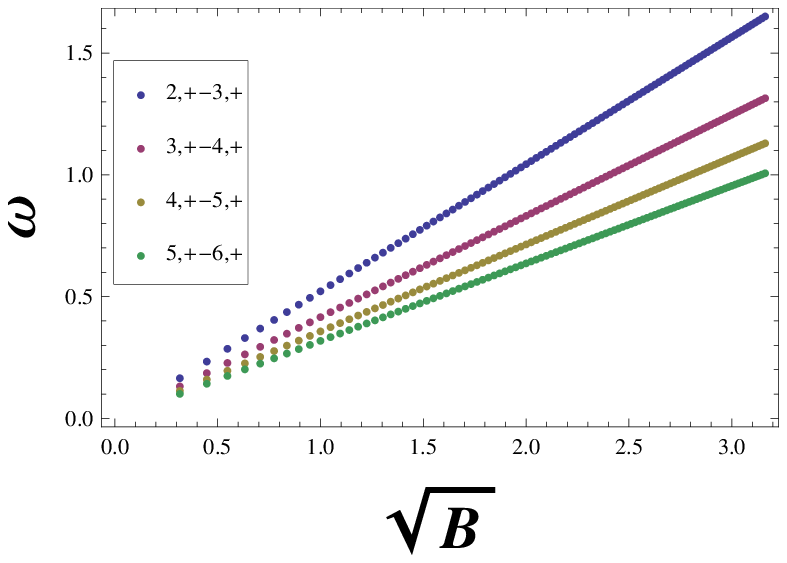}}
		\centerline{(b)}
	\end{minipage}
	\vfill
	\begin{minipage}{0.49\linewidth}
		\centerline{\includegraphics[width=1\textwidth]{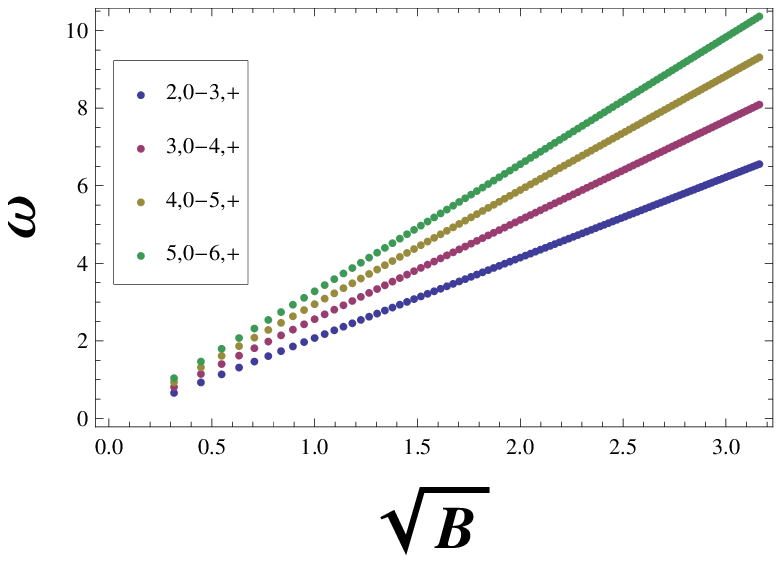}}
		\centerline{(c)}
	\end{minipage}
	\hfill
	\begin{minipage}{0.49\linewidth}
		\centerline{\includegraphics[width=1\textwidth]{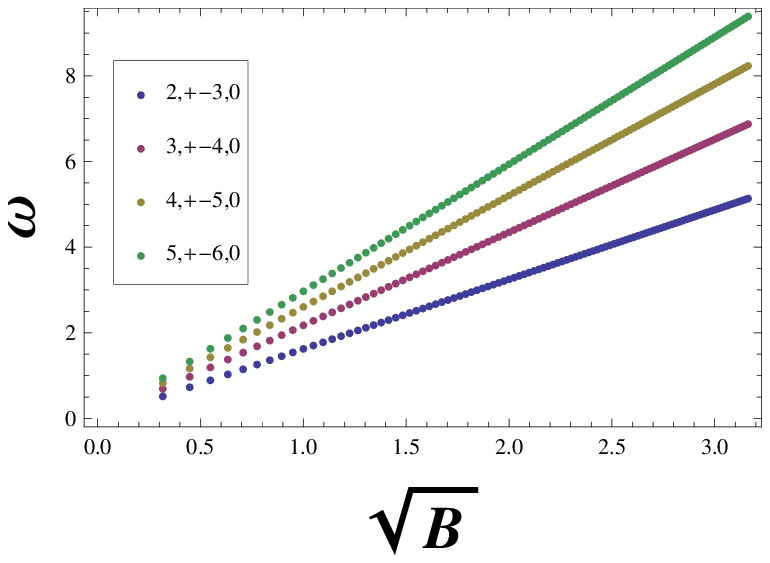}}
		\centerline{(d)}
	\end{minipage}
	\caption{ (color online) The linear behavior between the resonant peak $\omega$ and the square root of the external magnetic field $\sqrt{B}$ for $\mu=0$ for a photon-like fermion. (a) The peaks between $E_{n,0}$ and $E_{n+1,0}$ bands. (b) The peaks between $E_{n,+}$ and $E_{n+1,+}$ bands. (c) The peaks between $E_{n,0}$ and $E_{n+1,+}$ bands. (d) The peaks between $E_{n,+}$ and $E_{n+1,0}$ bands.} 
	\label{omega-b}
\end{figure}

{Furthermore,} from the numerical results (see Fig. \ref{omega-b}), the resonant peak frequency $\omega$ of photon-like fermion modes is linearly dependent on the square root of external magnetic field $B$ at zero chemical potential. In the case of Weyl fermion, this behavior is exactly related to the linear band structure of the Weyl fermion \cite{linear1,linear2,linear3}, while this may not be precise for photon-like fermion. The reason is that, for Weyl fermion, the gapped Landau levels are symmetric with respect to the $p_z=0$ axis  and hence the resonant transition occurs at $p_z=0$. The resonant frequency of Weyl fermion is proportional to the energy difference between two adjacent Landau levels at $p_z=0$, namely $\sqrt{B}$. This argument is not completely correct for photon-like fermion because the gapped Landau levels are slightly asymmetric to the $p_z=0$ axis (see Fig. \ref{fig1}). Therefore the resonant transition does not occur exactly at $p_z=0$,  but some other $p_z$ that extremizes the energy difference between the Landau levels allowed by the selection rules. This indicates that the linear dependence of $\sqrt{B}$ on the energy difference of Landau levels at $p_z=0$ is not rigorously related to the linear band structure of the photon-like fermion. 

\subsection{photon-like fermion gas}

In the Appendix \ref{app1}, we generalize the fermion doubling theorem \cite{NN} to the photon-like fermion gas, i.e., in a lattice model, the photon-like fermion nodes {with opposite chirality} emerge in pairs. All of the photon-like fermion pairs will contribute to the magneto-optical conductance. Notice that the Hamiltonian $H_B(p_z)$ has a quasi one-dimensional particle-hole symmetry. Namely, if we keep the two-dimensional Landau level structure unchanged, 
\begin{equation}
CH_B(p_z)C^\dagger=-H_B(-p_z),
\end{equation}
 where
\begin{equation}
C=
\left(
\begin{array}{ccc}
1
& 0
& 0 \\ 
0
& -1
& 0 \\ 
0
& 0
& 1
\end{array}
\right).
\end{equation}
Because of this symmetry,  the eigen wave function $\Psi_n^s=(a_n^s,b_n^s,c_n^s)^T$ of Hamiltonian $H_B$  in (\ref{hb}) pocesses the following property: If 
\begin{equation}
H_B(p_z)(a_n^s,b_n^s,c_n^s)^T=E_n^s(p_z)(a_n^s,b_n^s,c_n^s)^T,
\end{equation}
then,
\begin{eqnarray}
-H_B(-p_z)(a_n^s,-b_n^s,c_n^s)^T&=&E_n^{s}(p_z)(a_n^s,-b_n^s,c_n^s)^T,\nonumber\\
E_n^s(p_z)&=&-E_n^{-s}(-p_z),
\end{eqnarray} 
where $s$ is the band index. For later convenience, we denote Hamiltonian $H^L$ near one photon-like fermion node as $H^L=H_B$, then the Hamiltonian $H^R$ of the other photon-like fermion with opposite chirality reads: $H^R=-H^L$, we have
\begin{eqnarray}
E_n^{sL}(p_z)&=&-E_n^{-sR}(p_z), \label{prop1}\\ 
\Psi_n^{sL}(p_z)&=&\Psi_n^{-sR}(p_z).\label{prop2}
\end{eqnarray}
{Thus} 
\begin{equation}
\sigma_{\alpha\beta}^R=\sigma_{\alpha\beta}^L, \label{RL}
\end{equation}
where $\tilde{\Psi}_{ns}=(a_n^s,-b_n^s,c_n^s)$ (see Appendix \ref{appen1} for more details). Notice that we have used the fact that $\langle \tilde{\psi}|J_\alpha|\tilde{\psi}\rangle=\langle {\psi}|J_\alpha|{\psi}\rangle$, then $\sigma_{\alpha\beta}^R(\omega)$ is the same as $\sigma_{\alpha\beta}^L(\omega)$. Thus,  the magneto-optical conductance of  a pair of photon-like fermion nodes is the double of that of a single photon-like fermion node. This result then can be applied to the whole lattice system. The total magneto-optical conductance of a photon-like fermion gas is a simple summation of that from the contribution of each photon-like fermion node.

\section{Magneto-optical Conductance of Kane Fermion}

\subsection{Results for Kane Fermion}

We study the magneto-optical conductance of the gapless Kane fermion, which is the spinful photon-like fermion, similar to Weyl fermion versus gapless  Dirac fermion \cite{kane}. The effective Hamiltonian of a Kane fermion {under an external magnetic field} reads,
\begin{equation}
H_K=
\left(
\begin{array}{ccc}
H_B(p_z) & \delta \\ 
\delta^* & -H_B(p_z)
\end{array}
\right).
\end{equation}
$\delta$ is the mass matrix of Kane fermion. In the massless case, the effective Hamiltonian $H_K$ reduces to
\begin{equation}
H_K=H_\uparrow\oplus H_\downarrow,
\end{equation} 
with 
\begin{eqnarray}
H_\uparrow=H_B,\quad H_\downarrow=-H_B.
\end{eqnarray}
The current operator $J_i$ becomes,
\begin{equation}
J_i=\frac{\partial H_K}{\partial p_i}=J_i^\uparrow\oplus J_i^\downarrow,
\end{equation}
where 
\begin{equation}
J_i^\uparrow=j_i,\quad J_i^\downarrow=-j_i.
\end{equation}

One can easily check,
\begin{equation}
CH_{\uparrow,\downarrow}^B(p_z)C^\dagger=-H^B_{\uparrow,\downarrow}(-p_z)=H_{\downarrow,\uparrow}^B(-p_z),
\end{equation}
This means that the Hamiltonian $H_K$ is of a symmetry of the quasi-one dimensional particle-hole. 

The magneto-optical conductance reads,
	\begin{eqnarray}
	\sigma^K_{\alpha\beta}(\omega)&=&\frac{-i}{2\pi l_B^2}\sum_{n,n',s,s'}\int\frac{dk_z}{2\pi}\{\frac{f(E_n^{s\uparrow})-f(E_{n'}^{s'\uparrow})}{E_n^{s\uparrow}-E_{n'}^{s'\uparrow}}\nonumber\\
	&&\frac{\langle \Psi_{ns}^\uparrow|J_\alpha^\uparrow|\Psi_{n's'}^\uparrow\rangle\langle \Psi_{ns}^\uparrow|J_\beta^\uparrow|\Psi_{n's'}^\uparrow\rangle}{ \omega +E_n^{s\uparrow}-E_{n'}^{s'\uparrow}+i0^+}\nonumber\\
	&&+\frac{f(E_n^{s\downarrow})-f(E_{n'}^{s'\downarrow})}{E_n^{s\downarrow}-E_{n'}^{s'\downarrow}}\nonumber\\
	&&\frac{\langle \Psi_{ns}^\downarrow|J_\alpha^\downarrow|\Psi_{n's'}^\downarrow\rangle\langle \Psi_{ns}^\downarrow|J_\beta^\downarrow|\Psi_{n's'}^\downarrow\rangle}{ \omega +E_n^{s\downarrow}-E_{n'}^{s'\downarrow}+i0^+}\}\nonumber\\
	&=&\sigma^\uparrow+\sigma^\downarrow=2\sigma^\uparrow. \label{mokane}
	\end{eqnarray}

Because of the quasi one-dimensional particle-hole symmetry of $H_K(p_z)$, the magneto-optical conductance (\ref{mokane}) of a Kane node is the double of that of a photon-like fermion node (\ref{mofer}).

\begin{figure}
		\begin{minipage}{0.85\linewidth}
			\centerline{\includegraphics[width=1\textwidth]{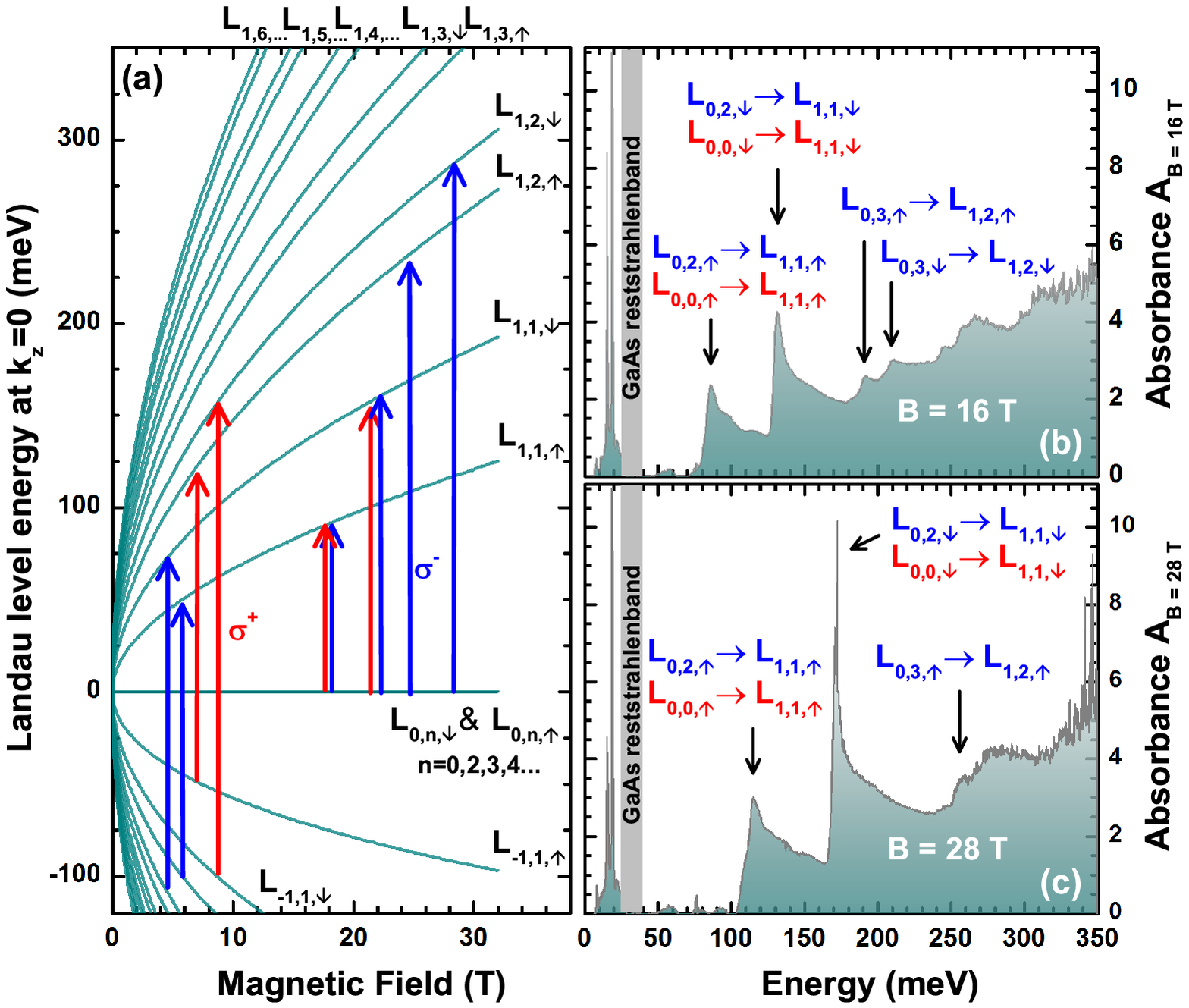}}
			\centerline{(a)}
		\end{minipage}\\
		\begin{minipage}{0.96\linewidth}
			\centerline{\includegraphics[width=1\textwidth]{T1-8mu02-2.eps}}
			\centerline{(b)}
		\end{minipage}
		\caption{
 			 (color online) (a) Landau levels (for $p_z=0$) and experimentally measured absorbance as a function of photon energy at a temperature of 1.8 K with $B=16T$ and $28T$ respectively in Hg$_{1-x}$Cd$_x$Te. (Adapted from Fig. 3 in Ref. \cite{kane}.)
			{(b) is the same as in Fig. \ref{xx}(d) for comparison with (a).}}
		\label{kane-xy}
	\end{figure}

\subsection{Comparing with experiments}

  The magneto-optical conductance of the Kane fermion was calculated in the literature \cite{mo-kane3,mo-kane}. However,  the low frequency  responses of the magneto-optics were ignored in previous calculations.  Experimentally, the effective Kane model of Cd$_3$As$_2$ is not valid down to arbitrarily low energies \cite{mo-kane2}. 
  
 For another candidate, Hg$_{1-x}$Cd$_x$Te, the magneto-optical absorbance was also measured. We adapt Fig. 3 in \cite{kane}  (see Fig. \ref{kane-xy}(a)) in order to compare with our study. {These data can be reasonably explained by our numerical calculation, especially those low frequency peaks} (see Fig. \ref{kane-xy}(b)). {Take the $\omega$=85meV, 130meV, 190meV and 210meV peaks at $B=16T$ in Fig. \ref{kane-xy}(a) as examples. The experimental magnetic length is $l_B(B=16T)=6.4\times10^{-9}m$. Similarly, in Fig. \ref{kane-xy}(b), we have three peaks around $\omega=0.94\hbar v_F/l_B$, $1.61\hbar v_F/l_B$, $2.06\hbar v_F/l_B$ and $2.17\hbar v_F/l_B$. There is only one tunable parameter, namely, the Fermi velocity. We can determine the Fermi velocity $v_F=0.880*10^6m/s$ by identify the $\omega=0.94\hbar v_F/l_B$ peak with the 85mev peak of the experimental data, then the other three peaks in Fig. \ref{kane-xy}(b) correspond to 146mev, 188meV and 198meV. There are other two peaks at $0.52\hbar v_F/l_B$=47meV and $1.11\hbar v_F/l_B$=100meV which may correspond to the bumps in Fig. \ref{kane-xy}(a) around 55meV and 98meV. These bumps are not explained in Ref. \cite{kane}. The two peaks near zero frequency corresponds to 3meV and 7meV where the corresponding peaks in \ref{kane-xy}(a) were explained as phonon contribution \cite{kane}. In other words, these peaks near zero frequency can be explained without involving external phonons. To sum, our results are in agreement with the experimental data qualitatively, and we can also provide reasonable explanations for the peaks that were not explained in Ref. \cite{kane}.}

\subsection{Disorder effect}		
{The low energy gapless chiral modes are stable against non-magnetic weak disorders. For simplicity, we assume the disorder does not mix the states in different Landau levels, and only consider the zero mean random Gaussian scalar disorder $A(z)$ with correlation reads \cite{dis1,dis2}, 
	\begin{equation}
	\langle A(z)A(z')\rangle=\frac{\Delta_D}{l_m}\delta(z-z'),
	\end{equation}
	where $\Delta_D$ is the dimensionless disorder strength, and $l_m$ is the mean free path. We also do not consider the rare region effects for simplicity\cite{rare}. Consider the effective Hamiltonian in the small $p_z$ limit near a node. For $n=0$ or a given $n>1$,
	\begin{equation}
	H=H_{low}+H',
	\end{equation}
	with $H_{low}=v_F \hbar p_z\sigma_z$ for the 0th Landau level and $H_{low}=-\sqrt{2n-1}\hbar v_F p_z \sigma_z$ (see, Eq. (\ref{low})) being the effective Hamiltonian for the $n$th gapless Landau excitation,
	and
	\begin{equation}
	H'=v_D\int d^2x \Psi^\dagger(x)A(x)\Psi(x).
	\end{equation}
	To consider the disorder effects, we derive the renormalization group flow equation by averaging over the disorder potential through the replica method \cite{dis3}. After integrating the disorder field $A(x)$, the effective action reads,
	\begin{eqnarray}
	S&=&\sum_{\alpha}\int \frac{d\omega dp_z}{(2\pi)^4}\Psi^\dagger_\alpha(i\omega,p_z)[i\omega-H_{low}]\Psi_\alpha(i\omega, p_z)\nonumber\\
	&&+\frac{\Delta}{2l_m}\int \frac{d\omega_1 d\omega_2dp_{z1} dp_{z2} dp_{z3}}{(2\pi)^{5}}\Psi_\alpha^\dagger(i\omega_1,p_{z1})\Psi_\alpha(i\omega_1,p_{z2})\nonumber\\
	&&\Psi_\beta^\dagger(i\omega_2,p_{z3})\Psi_\beta(i\omega_2,p_{z1}+p_{z2}+p_{z3}),
	\end{eqnarray}
	with $\Delta=\Delta_D v_D^2$. We integrate over the momentum shell of the fast moving field to calculate the one-loop diagrams of self energy and the vertex correction.\\ }

\begin{figure}[ptb]
 	\centering
 	\includegraphics[width=0.3\textwidth]{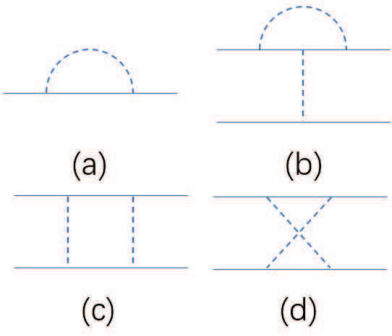}\newline\caption{  (color online) The sketch of Feynman diagrams of one-loop corrections to self energy (a) and the disorder strength (b,c,d).}
 	\label{feynman}                                                                 
 \end{figure}

{(1) For the leading order of loop correction of self energy, we calculate the loop diagram in Fig. \ref{feynman}(a).
	\begin{eqnarray}
	\Sigma_{dis}(i\omega)&=&\frac{\Delta}{l_m}\int^{\Lambda}_{\Lambda e^{-l}}\frac{dp_z}{2\pi}TrG_0(i\omega,p_z)\nonumber\\
	&=&\frac{\Delta}{l_m}\int^{\Lambda}_{\Lambda e^{-l}}\frac{dp_z}{2\pi}(-\frac{2 i \omega}{{p_z}^2+\omega^2})\nonumber\\
	&=&\frac{\Delta}{\pi l_m}(-\frac{ i \Lambda l \omega}{\Lambda^2+\omega^2}), 
	\end{eqnarray}
	where $G_0=\frac{1}{i\omega-H_{low}}$ being the free Green's function, and $\Lambda$ being the cut-off, which is of the order of the inverse of the lattice constant. \\}
		
{(2) Leading order of vertex correction of Fig. \ref{feynman}(b),
	\begin{eqnarray}
	\delta\Delta_{dis}&=&\frac{\Delta^2}{l_m}\int\frac{dp_z}{2\pi}TrG_0(p_z)^2\nonumber\\
	&=&\frac{\Delta^2}{l_m}\int\frac{dp_z}{2\pi}(\frac{2 \left(p_z^2-\omega^2\right)}{\left({p_z}^2+\omega^2\right)^2})\nonumber\\
	&=&\frac{\Delta^2}{\pi l_m}(\frac{ \Lambda l \left(\Lambda^2-\omega^2\right)}{\left(\Lambda^2+\omega^2\right)^2}).
	\end{eqnarray}
	The vertex contributions of Fig. \ref{feynman}(c) and \ref{feynman}(d) cancel each other.}
	
{To sum, as long as the mean free path $l_m$ is large enough, in the $\Lambda\rightarrow \infty$ limit, the RG equations are
	\begin{equation}
	\frac{dv_F}{dl}=0, \quad \frac{d\Delta}{dl}=0.
	\end{equation}
    This shows that the disorder is an irrelevant perturbation to the low energy Landau level structure of the Kane fermion. Notice that for the free Kane fermion without magnetic field, the weak disorder is a relevant perturbation and drives the system into a diffusive state controlled by disorder, similar to the case of Weyl fermion \cite{dis1,dis4}. The existence of the magnetic field suppresses the influence of disorder.  }

\section{quadratic correction}
We may read out more informations from the $E_n^0$ bands by adding a quadratic term {$\frac{p_i^2}{2m^*}$ in the Hamiltonian $H_B$, where $m^*$ is the effective band mass.} This quadratic term will break the quasi one-dimensional particle-hole symmetry of $H_0$ and lift the degeneracy at $p_z=0$ of $E_n^0$ Landau levels. The Hamiltonian $H_Q$ under a constant magnetic field $B$ becomes \cite{photon-like fermion},
\begin{equation}
H_{QB}=H_B(p_z)+\frac{1}{2m^*}(p_z^2+2Ba^\dagger a+B),
\label{hqu}
\end{equation}
where $H_B(p_z)$ is the Hamiltonian (\ref{hb}). 

\begin{figure}[ptb]
	\centering
	\includegraphics[width=0.3\textwidth]{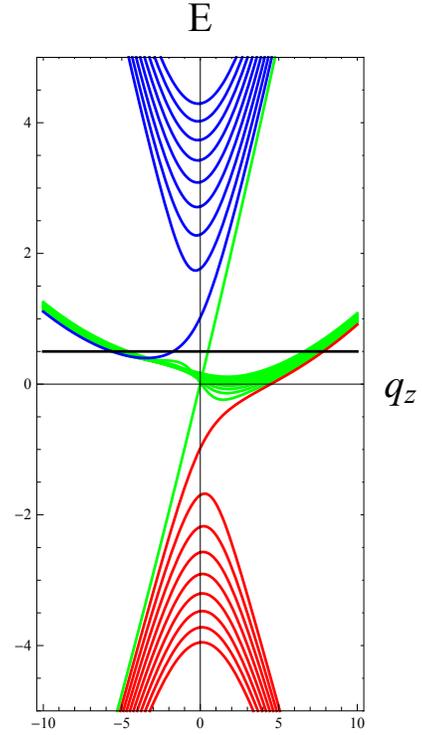}\newline\caption{  (color online) A sketch diagram of Landau Level structure for the  photon-like fermion with quadratic corrections (\ref{hqu}) where $l_B=1$ and {$m^*=0.02/l_B$}. The meaning of the colored curves is similar to that in Fig.\ref{fig1}. The absorptions in the conductance are plotted in Fig. \ref{xx-xy-m}.}
	\label{qua}
\end{figure}

For a given eigen wave function of $H_B$, for instance, $\Psi(n)$,
\begin{equation}
 H_{QB}(p_z)\Psi(n)=[H_B(p_z-\omega_c)+\frac{p_z^2}{2m^*}+\omega_c(n+\frac{1}{2})]\Psi(n), 
\end{equation}
with $\omega_c=B/m^*$. This relation shows that, the quadratic correction does not change the eigen wave functions and the eigen energy satisfies
\begin{equation}
E_{QB}(p_z,n,s)=E_B(p_z-\omega_c,n,s)+\frac{p_z^2}{2m^*}+\omega_c(n+\frac{1}{2}),\label{spec-qu}
\end{equation}
where $n$ is the Landau level index, and $s=0,\pm$ labels the bands within the $n$th Landau level. An example of the quadratic correction to the Landau level spectrum is shown in Fig. \ref{qua}. 

\begin{figure}
	\begin{minipage}{0.85\linewidth}
		\centerline{\includegraphics[width=1\textwidth]{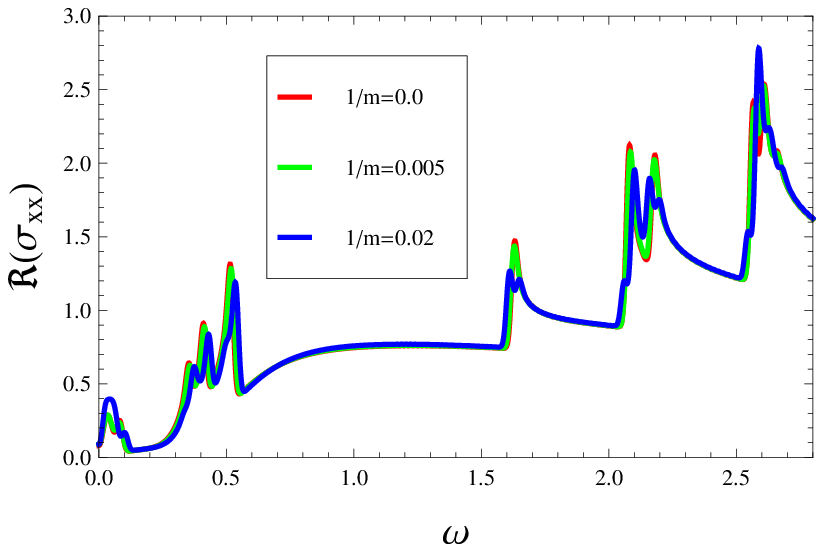}}
		\centerline{(a)}
	\end{minipage}\\
	\begin{minipage}{0.8\linewidth}
		\centerline{\includegraphics[width=1\textwidth]{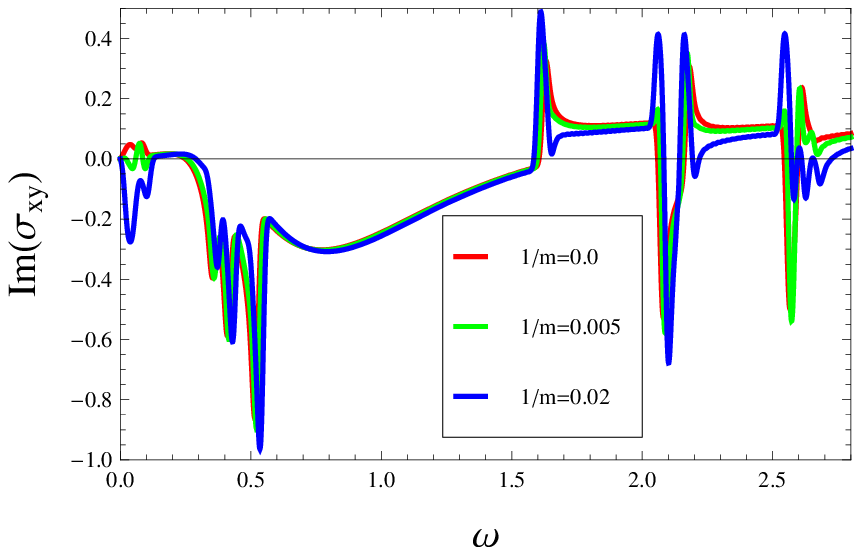}}
		\centerline{(b)}
	\end{minipage}
	\caption{ (color online) (a) and (b) are the real part of $\sigma_{xx}$ and the imaginary part of $\sigma_{xy}$ with quadratic corrected photon-like fermion. The masses are {$1/m=0$, $1/m=0.005l_B$, and $1/m=0.02l_B$} respectively. {$T=0.5/l_b$ and $\mu=0.01/l_B$}. The peaks near zero frequency are magnificently enhanced by the quadratic terms. {Especially, the ones in (b) are tuned from positive to negative.} The other high energy peaks are also split by the correction term.}
	\label{xx-xy-m}
\end{figure}

Furthermore, because of the quadratic correction, the currents $j^Q_\alpha=\frac{\partial H^Q_B}{\partial p_\alpha}$ become
\begin{eqnarray}
j^Q_x&=&j_x+\frac{1}{m^*}\sqrt{\frac{B}{2}}(a+a^\dagger),\\
j^Q_y&=&j_y-\frac{i}{m^*}\sqrt{\frac{B}{2}}(a-a^\dagger).
\end{eqnarray}
Therefore the selection rules remain unchanged, namely, $n\rightarrow n\pm 1$. From the spectrum relation (\ref{spec-qu}) and the unchanged selection rules, the resonant peaks will split to $\omega\rightarrow \omega\pm\omega_c$. 

We plot the magneto-optical conductance with quadratic correction to the spectrum in Fig. \ref{xx-xy-m}. We see that the mass term flattens the extrema of Landau levels, the absorption strength becomes stronger because the resonant frequency is determined by the difference between the extrema and the density of states becomes larger in the presence of the mass term. For a large enough mass $m^*$, the peak splitting becomes obvious. {For example, see the blue curve in Fig. \ref{xx-xy-m}(a). The magnitude of the peak splitting is $2\omega_c$.} Furthermore, the quadratic term lifts the infinite degeneracy at $p_z=0$ of the gapless Landau levels.  The peaks near the zero frequency is enhanced magnificently. This fact provides a strong evidence for the existence of the gapless Landau levels instead of exact flat bands, which is experimentally detectable.

\section{Conclusions}

We studied the magneto-optical conductance of the photon-like fermion and the Kane fermion. {The resonant peaks show a strong dependence on chemical potential at low temperature.} Besides the ordinary resonant peaks as in Weyl fermion, there are extra peaks stemming from the gapless Landau levels which are absent in the Weyl fermion case. {Especially, there are peaks near zero frequency coming from the transition between the gapless Landau modes, and splitting of the peaks  stemming from the transition between the gapless Landau levels and ordinary gapped ones, both of which show strong evidences for the existence of gapless modes instead of flat bands.} {We show that the low energy gapless Landau levels in the Kane fermion are stable to weak disorders.} We also considered the effects of quadratic correction. These phenomena would be experimentally detected in photon-like fermion material candidates. Applying our calculation to the Kane fermion, we gave the magneto-optical absorbance measurements for Hg$_{1-x}$Cd$_x$Te in low frequencies a reasonable explanation.

\acknowledgments

We thank Bin Chen, Qian Niu and Yong-Shi Wu for helpful discussions. This work is supported by NNSF of China with No. 11474061 (XL, YGC, YY) and No. 11804223 (XL).\\

\clearpage

\appendix

\section{Proof of Eq. (\ref{RL})} \label{appen1}

Using the properties Eq. (\ref{prop1}) and (\ref{prop2}), the magneto-optical conductance $\sigma_{\alpha\beta}^R$ can be rewritten as

\begin{widetext}
	\begin{eqnarray*}
		\sigma_{\alpha\beta}^R&=&\int_{-\infty}^\infty\frac{dk_z}{2\pi}\frac{f(E_n^{sR}(k_z))-f(E_{n'}^{s'R}(k_z))}{E_n^{sR}(k_z)-E_{n'}^{s'R}(k_z)}\frac{\langle \Psi_{ns}^R(k_z)|J_\alpha^R|\Psi_{n's'}^R(k_z)\rangle\langle \Psi_{ns}^R(k_z)|J_\beta^R|\Psi_{n's'}^R(k_z)\rangle}{ \omega +E_{ns}^R(k_z)-E_{n's'}^R(k_z)+i0^+},\\
		&=&-\int\frac{d(-k_z)}{2\pi}\frac{f(E_n^{sR}(k_z))-f(E_{n'}^{s'R}(k_z))}{E_n^{sR}(k_z)-E_{n'}^{s'R}(k_z)}\frac{\langle \Psi_{ns}^R(k_z)|J_\alpha^R|\Psi_{n's'}^R(k_z)\rangle\langle \Psi_{ns}^R(k_z)|J_\beta^R|\Psi_{n's'}^R(k_z)\rangle}{ \omega +E_{ns}^R(k_z)-E_{n's'}^R(k_z)+i0^+}\\
		&=&-\int_{\infty}^{-\infty}\frac{dp_z}{2\pi}\frac{f(E_n^{sR}(-p_z))-f(E_{n'}^{s'R}(-p_z))}{E_n^{sR}(-p_z)-E_{n'}^{s'R}(-p_z)}\frac{\langle \Psi_{ns}^R(-p_z)|J_\alpha^R|\Psi_{n's'}^R(-p_z)\rangle\langle \Psi_{ns}^R(-p_z)|J_\beta^R|\Psi_{n's'}^R(-p_z)\rangle}{ \omega +E_{ns}^R(-p_z)-E_{n's'}^R(-p_z)+i0^+}\\
		&=&-\int\frac{dp_z}{2\pi}\frac{f(-E_n^{-sR}(p_z))-f(-E_{n'}^{-s'R}(p_z))}{-E_n^{-sR}(p_z)+E_{n'}^{-s'R}(p_z)}\frac{\langle \Psi_{ns}^R(-p_z)|J_\alpha^R|\Psi_{n's'}^R(-p_z)\rangle\langle \Psi_{ns}^R(-p_z)|J_\beta^R|\Psi_{n's'}^R(-p_z)\rangle}{ \omega -E_{n-s}^R(p_z)+E_{n'-s'}^R(p_z)+i0^+}\\
		&=&-\int_{\infty}^{-\infty}\frac{dp_z}{2\pi}\frac{f(E_n^{sL}(p_z))-f(E_{n'}^{s'L}(p_z))}{E_n^{sL}(p_z)-E_{n'}^{s'L}(p_z)}\frac{\langle \tilde{\Psi}_{ns}^L(p_z)|J_\alpha^L|\tilde{\Psi}_{n's'}^L(p_z)\rangle\langle \tilde{\Psi}_{ns}^L(p_z)|J_\beta^L|\tilde{\Psi}_{n's'}^L(p_z)\rangle}{ \omega +E_{ns}^L(p_z)-E_{n's'}^L(p_z)+i0^+}\\
		&=&\int_{-\infty}^{\infty}\frac{dp_z}{2\pi}\frac{f(E_n^{sL}(p_z))-f(E_{n'}^{s'L}(p_z))}{E_n^{sL}(p_z)-E_{n'}^{s'L}(p_z)}\frac{\langle \tilde{\Psi}_{ns}^L(p_z)|J_\alpha^L|\tilde{\Psi}_{n's'}^L(p_z)\rangle\langle \tilde{\Psi}_{ns}^L(p_z)|J_\beta^L|\tilde{\Psi}_{n's'}^L(p_z)\rangle}{ \omega +E_{ns}^L(p_z)-E_{n's'}^L(p_z)+i0^+}=\sigma_{\alpha\beta}^L.
	\end{eqnarray*}
\end{widetext}

\section{Generalization of Nielsen-Ninomiya no-go theorem to the photon-like fermion semimetals}\label{app1}

In 1981, Nielsen and Ninomiya published a series of papers proving a no-go theorem which stated the number of Weyl points with opposite chirality should be the same on a lattice satisfying four assumptions \cite{NN}:

(i) The Hamiltonian is local.

(ii) The lattice is translational invariant.

(iii) The Hamiltonian is Hermitian.

(iv) There is at least one quantized conserved charge Q for the fermion field.

In the following, we generalize Nielsen-Ninomiya no-go theorem to the photon-like fermion semimetals with an additional assumption, namely, only photon-like fermion points are involved in the lattice model near the Fermi surface.
\\

Near a photon-like fermion point, the Hamiltonian can be expanded as,
\begin{equation}
H^{(3)}=p_aV^a_iS^i + \mathcal{O}(p^2), \label{a1}
\end{equation}
where $(S^i)_{jk}=-i\epsilon_{ijk}$ being the spin-1 representation of the generators of the angular momentum algebra. The photon-like fermion point is right(left) handed if $\det V>0(\det V<0)$. Define $P_i\equiv p_aV^a_i$, then the spectra and eigen wavefunctions in spherical coordinates read,
\begin{eqnarray}
E_0&=&0, \quad \psi_0=( \sin \theta\cos \phi,\sin \theta \sin\phi,\cos\theta)^T,\nonumber\\
E_+&=&|{\bf{P}}|,\nonumber\\
 \psi_+&=&(\frac{-\cos\theta\cos\phi+i\sin\phi}{\sqrt{2}},\frac{-i\cos\phi-\cos\theta\sin\phi}{\sqrt{2}},\frac{\sin\theta}{\sqrt{2}})^T,\nonumber\\
E_-&=&-|{\bf{P}}|,\nonumber\\
 \psi_-&=&(\frac{-\cos\theta\cos\phi-i\sin\phi}{\sqrt{2}},\frac{i\cos\phi-\cos\theta\sin\phi}{\sqrt{2}},\frac{\sin\theta}{\sqrt{2}})^T.\nonumber\\
\end{eqnarray} 
As we discussed in our former paper on the photon-like fermion semimetal \cite{photon-like fermion}, for the free Hamiltonian, the zero energy flat band $E_0$ generates an emergent gauge symmetry. The reason is that, from the equation of motion of Hamiltonian (\ref{a1}),
\begin{equation}
i\partial_t\psi_j=-i\partial_iS^i_{jk}\psi_k.
\end{equation}
Since $S^i_{jk}$ is totally anti-symmetric, then, multiplying both sides with $\partial_j$,
\begin{equation}
i\partial_t(\partial_j\psi_j)=0,
\end{equation}
which means $\partial_j\psi_j=C(x_i)$. Notice that there is a zero energy wave function $\psi_0=\partial_i\Lambda(x_i)$ for the  Hamiltonian (\ref{a1}), and if we choose $\Lambda(x_i)$ to satisfy $\partial^2\Lambda(x_i)=C(x_i)$, then we can consider a new state, 
\begin{equation}
\psi_j=\psi_j'+\partial_j\Lambda(x_i),
\end{equation} 
and,
\begin{equation}
\partial_j\psi_j'=0,
\end{equation}
which is a constraint on the wave function $\psi'_j$, namely, only two of the components in $\psi'_j$ are independent. Therefore the flat band is a redundant gauge degree of freedom and unphysical which should be projected out, which is similar to the unphysical longitudinal photon \cite{photon-like fermion}. Later we will also prove that the flat band wave function is topologically trivial and will not contribute to the topological number which is associated with the chirality of the photon-like fermion point, namely, the exact flat band will not affect the no-go theorem. 

In a proper basis, the Taylor expansion of the N-band lattice Hamiltonian near the photon-like fermion point reads \cite{NN},
\begin{eqnarray}
H^{(3)}_N&\sim&
p_a V^a_i  
\left(
\begin{array}{cccccc}
 & (i-1) & (i) & (i+1) & & \\
{\bf {0}} & & & & & \\ 
 & S^i_{11} & S^i_{12} & S^i_{13} & & (i-1) \\
& S^i_{21} & S^i_{22} & S^i_{23} & &  (i)\\
& S^i_{31} & S^i_{32} & S^i_{33} & &  (i+1)\\
& & & &{\bf {0}} &
\end{array}
\right)\nonumber\\
&&+p_a
\left(
\begin{array}{ccccc}
	b_1 & & & & 0 \\ 
	& \cdot &  &  &  \\
	&  & \cdot &  & \\
	&  &  & \cdot & \\
	0 & & & & b_N 
\end{array}
\right)\nonumber\\
&&+p_a
\left(
\begin{array}{cccccc}
& (i-1) & (i) & (i+1) & & \\
{\bf {0}} & *& *& *& * & \\ 
*& 0 & 0 & 0 & *& (i-1) \\
*& 0 & 0 & 0 & *&  (i)\\
*& 0 & 0 & 0 & *&  (i+1)\\
* & *& *& *&{\bf {0}} &
\end{array}
\right)\nonumber\\
&&+\mathcal{O}(p^2),
\end{eqnarray}
where $(i-1)$, $(i)$, and $(i+1)$ label the $(i-1)$-th, $(i)$-th, and $(i+1)$-th bands which converge at the photon-like fermion point. To the lowest order, the eigen wave function is determined by the first term. For instance, the energy of the $(i-1)$-th positive energy band reads $|{\bf p} |$ with the N-dimensional wave function being,
\begin{equation}
|\omega_+(p_i)\rangle =
\left(
\begin{array}{ccccc}
{\bf {0}} \\ 
\psi_+^1  \\
\psi_+^2 \\
\psi_+^3 \\
{\bf {0}} 
\end{array}
\right)
\begin{array}{ccccc}
 \\ 
i-1  \\
i \\
i+1 \\
\\
\end{array}.
\end{equation}
Now we define a normalized function 
\begin{equation}
f_+(\theta,\phi)=|\omega_+(\theta,\phi)\rangle
\end{equation}
where the domain of $f_+$ is on an infinitesimal sphere $S^2$ around the photon-like fermion point. The map $f_+$ is an element in the homotopy group $\pi_2(CP^{N-1})$. The map $f_+(\theta,\phi)$ restricted at the south pole $\theta=\pi$ reads,
\begin{equation}
f_+(\pi,\phi)=e^{i\phi}
\left(
\begin{array}{ccccc}
{\bf {0}} \\ 
1/\sqrt{2}  \\
-i/\sqrt{2} \\
0 \\
{\bf {0}} 
\end{array}
\right)
\begin{array}{ccccc}
\\ 
i-1  \\
i \\
i+1 \\
\\
\end{array}.
\end{equation}
Notice that the south pole on $S^2$ can be viewed as the whole boundary of $E^2$ mapping to the same point, therefore the class $[f|_{S^1}]\in \pi_1(s^1)$ is the winding number which is $+1$ or $-1$. When $\det V>0$, the coordinate system near the photon-like fermion point is right-handed, then the winding number is $+1$, and when $\det V<0$, the coordinate system near photon-like fermion point is left-handed and the winding number is $-1$. Similarly, for the negative energy branch, when $\det V>0$, the winding number is $-1$, and when $\det V<0$, the winding number is $+1$. For the exact flat band, the winding number is zero because $f_0(\pi,\theta)=(0,0,...0,0,-1,0....0)^T$. To sum, for the infinitesimal $S^2$ sphere surrounding a degenerate photon-like fermion point, the positive(negative) energy branch with positive(negative) helicity corresponds to $+1$ element of $\pi_2(CP^{N-1})$, and the positive(negative) energy branch with negative(positive) helicity corresponds to $-1$ element.

Because of the periodicity of the Brillouin zone and the additivity of the $\pi_2(CP^{N-1})$ group, we have \cite{NN}
\begin{equation}
[\hat{f}_{BS}]=\sum_i[\hat{f}_i]=0, \label{sum}
\end{equation}
where $\hat{f}_{BS}$ imbeds the Brillouin zone surface $S^2$ into $CP^{N-1}$, $[\hat{f}_{BS}]$ denotes the corresponding element in $\pi_2(CP^{N-1})$, and the summation $i$ runs over all the degenerate points $i$. The Eq. (\ref{sum}) means the total winding number for the degenerate points between the $i$-th and $(i+1)$-th or $(i-1)$-th bands are zero (we have omitted the exact flat bands and assumed only photon-like fermion points are involved), namely,
\begin{equation}
N_r(i,i+1)-N_r(i-1,i)=N_l(i,i+1)-N_l(i-1,i),
\end{equation}
where $N_r(i+1,i)$ is the number of degenerate points between $i$-th and $(i+1)$-th bands with the upper $i$-th having positive helicity. Notice that for the highest band, $N_r(0,1)=N_l(0,1)=0$, therefore we have,
\begin{equation}
N_r(i,i+1)=N_l(i,i+1). \label{NN-f}
\end{equation}
The Eq. (\ref{NN-f}) proves our generalization of the Nielsen-Ninomiya no-go theorem, namely there are equal number of the left-handed photon-like fermion points and the right-handed ones in a lattice model.


\begin{thebibliography}{99}
	
	
	\bibitem{weyl1} Xi. G. Wan, A. M. Turner, A. Vishwanath, and S. Y. Savrasov, Phys. Rev. B {\bf 83}, 205101 (2011).
	
	\bibitem{weyl2} S.-M. Huang, S.-Y. Xu, I. Belopolski, C.-C. Lee, G. Chang, B. K.  Wang, N.  Alidoust, G. Bian, M.  Neupane,  C. Zhang, S.  Jia, A.  Bansil, H. Lin, and M. Z.  Hasan, Nat. Commun. {\bf 6}, 7373 (2015).
	
	\bibitem{weyl3} H. M.  Weng, C. Fang, Z. Fang, B. A. Bernevig, and Xi. Dai, Phys. Rev. X {\bf 5}, 011029 (2015).
	
	\bibitem{dirac1} Z. J. Wang, H. M. Weng, Q. S. Wu, X. Dai, Z. Fang, Phys. Rev. B {\bf 88}, 125427 (2013).
	
	
	
	\bibitem{dirac2} M. Neupane, S. Y. Xu, R. Sankar, N. Alidoust, G. Bian, C. Liu, I. Belopolski, T.-R. Chang, H.-T. Jeng, H. Lin, A. Bansil, F. C. Chou, M. Z. Hasan, Nature Commun. {\bf 05}, 3786 (2014).
	
	\bibitem{dirac3} S. Borisenko, Q. Gibson, D. Evtushinsky, V. Zabolotnyy, B. Buechner, R. J. Cava, Phys. Rev. Lett. {\bf 113}, 027603 (2014).
	
	\bibitem{dirac4} Z. K. Liu, B. Zhou, Z. J. Wang, H. M. Weng, D. Prabhakaran, S. -K. Mo, Y. Zhang, Z. X. Shen, Z. Fang, X. Dai, Z. Hussain, Y. L. Chen, Science {\bf 343}, 864 (2014).
	
	
	\bibitem{burkov2014} A. A. Burkov, Phys. Rev. Lett. {\bf 113}, 187202 (2014).
	
	\bibitem{weyl-rmp} N. P.  Armitage, E. J. Mele, and A. Vishwanath, Rev. Mod. Phys. {\bf 90}, 15001 (2018).
	
	\bibitem{linear1} V. P. Gusynin, S. G. Sharapov, and J. P. Carbotte, Phys. Rev. Lett. {\bf 98}, 157402 (2007).
	
	\bibitem{linear2} P. E. C. Ashby and J. P. Carbotte, Phys. Rev. B {\bf 87}, 245131 (2013).
	
	\bibitem{linear3} J. D. Malcolm and E. J. Nicol, Phys. Rev. B {\bf 90}, 035405 (2014).
	
	\bibitem{kane} M. Orlita, D. M. Basko, M. S. Zholudev, F. Teppe, W. Knap, V. I. Gavrilenko, N. N. Mikhailov, S. A. Dvoretskii, P. Neugebauer, C. Faugeras, A.-L. Barra, G. Martinez, and M. Potemski,  Nature Physics {\bf 10}, 233 (2014).
	
	\bibitem{mo-kane2} A. Akrap, M. Hakl, S. Tchoumakov, I. Crassee, J. Kuba, M. O. Goerbig, C. C. Homes, O. Caha, J. Novak, F. Teppe, W. Desrat, S. Koohpayeh, L. Wu, N. P. Armitage, A. Nateprov, E. Arushanov, Q. D. Gibson, R. J. Cava, D. van der Marel, B. A. Piot, C. Faugeras, G. Martinez, M. Potemski, and M. Orlita, Phys. Rev. Lett. {\bf 117}, 136401 (2016).
	\bibitem{type2} A. A. Soluyanov, D. Gresch, Z. J.  Wang, Q. S. Wu, M. Troyer, X. Dai, and B. A. Bernevig, Nature {\bf 527}, 495 (2015).
	
	\bibitem{new} B. Bradlyn, J. Cano, Z. J.  Wang, M. G. Vergniory, C. Felser, R. J. Cava, and B. A. Bernevig, Science {\bf aaf5037} (2016).
	
	
	\bibitem{hourglass} Z. J.  Wang, A.  Alexandradinata, R. J. Cava, and B. A. Bernevig, Nature {\bf 532}, 189 (2016).
	
	\bibitem{tri} B. Q. Lv, Z.-L. Feng, Q.-N. Xu, J.-Z. Ma, L.-Y. Kong,
	P. Richard, Y.-B. Huang, V. N. Strocov, C. Fang, H.-M.
	Weng, Y.-G. Shi, T. Qian, and H. Ding, Nature {\bf 546}, 627
	(2017).
	
	\bibitem{tri2} K.-H. Ahn, W. E. Pickett, K.-W. Lee, Phys. Rev. B {\bf 98}, 035130 (2018)
	
	\bibitem{tri3} H. Yang, J. B.  Yu, S. S. P. Parkin, C. Felser, C.-X. Liu, B.  H. Yan, Phys. Rev. Lett. {\bf 119}, 136401 (2017).
	
	\bibitem{tri4} Z. M. Zhu, G. W. Winkler, Q. S. Wu, J. Li, A. A. Soluyanov, Phys. Rev. X {\bf 6}, 031003 (2016).
	
	\bibitem{sym1} H. C. Po, A. Vishwanath, H. Watanabe, Nat. Commun. {\bf 8}, 50 (2017).
	
	\bibitem{sym2} F. Tang, H. C. Po, A. Vishwanath, and X. G.  Wan, Nature Physics {\bf 15}, 470 (2019), Science Advances {\bf 5}, eaau8725 (2019), Nature {\bf 566}, 486 (2019).
	
	\bibitem{sym3} Barry Bradlyn, L. Elcoro, Jennifer Cano, M. G. Vergniory, Zhijun Wang, C. Felser, M. I. Aroyo, and B. Andrei Bernevig, Nature {\bf 547}, 298 (2017). 
	
	\bibitem{extra1} T. Morimoto, A. Furusaki, Phys. Rev. B {\bf 89}, 235127 (2014).
	
	\bibitem{extra2} M. Kargarian, M. Randeria, Y.-M. Lu, PNAS {\bf 113}, 8648 (2016).
	
	
	
	
	\bibitem{photon-like fermion} X. Luo, F.-Y. Li, and Y. Yu, New J. Phys., {\bf 20} 083036 (2018).
	
	\bibitem{mo-kane3} J. D. Malcolm and E. J. Nicol, Phys. Rev. B {\bf 92}, 035118, (2015)
	
	\bibitem{mo-kane} J.  D. Malcolm and E, J. Nicol, Phys. Rev. B {\bf 94} 224305 (2016).
	
	
	
	\bibitem{lieb} E. H. Lieb, Phys. Rev. Lett. {\bf 62}, 1201 (1989).
	
	
	\bibitem{mahan} See, e.g. G. D. Mahan, Many-Particle Physics, (Springer, 2000).
	
	
	\bibitem{NN} H. B. Nielsen and M. Ninomiya, Nucl. Phys. {\bf B185}, 20 (1981); {\it ibid}, Nucl. Phys. {\bf B193}, 173 (1981);  {\it ibid},Phys. Lett. B {\bf 130}, 389 (1983). 
	
	
	
	
	
	
	\bibitem{dis1} Jing Wang, Peng-Lu Zhao, Jing-Rong Wang, and Guo-Zhu Liu, Phys. Rev. B {\bf 95}, 054507 (2017).
	
	\bibitem{dis2} Björn Sbierski, Kevin S.C. Decker, Piet W. Brouwer, Phys. Rev. B {\bf 94}, 220202 (2016).
	
	\bibitem{rare} Rahul Nandkishore, David A. Huse, and S. L. Sondhi, Phys. Rev. B {\bf 89}, 245110 (2014).
	
	
	\bibitem{dis3} P. Goswami and S. Chakravarty, Phys. Rev. Lett. {\bf 107}, 196803 (2011).
	
	\bibitem{dis4} A. Altland, B. D. Simons, and M. R. Zirnbauer, Phys. Rep. {\bf 359}, 283 (2002).
	
\end{thebibliography}
\end{document}